# Analyzing Web Application Log Files to Find Hit Count Through the Utilization of Hadoop MapReduce in Cloud Computing Environment

Sayalee Narkhede
Department of Information Technology
Maharashtra Institute of Technology
Pune, India
sayleenarkhede@gmail.com

Trupti Baraskar
Department of Information Technology
Maharashtra Institute of Technology
Pune, India
baraskartn@gmail.com

Debajyoti Mukhopadhyay
Department of Information Technology
Maharashtra Institute of Technology
Pune, India
debajyoti.mukhopadhyay@gmail.com

*Abstract*—MapReduce has been widely applied in various fields of data and compute intensive applications and also it is important programming model for cloud computing. Hadoop is an open-source implementation of MapReduce which operates on terabytes of data using commodity hardware. We have applied this Hadoop MapReduce programming model for analyzing web log files so that we could get hit count of specific web application. This system uses Hadoop file system to store log file and results are evaluated using Map and Reduce function. Experimental results show hit count for each field in log file. Also due to MapReduce runtime parallelization response time is reduced.

## I. INTRODUCTION

Recently the computing world has been undergoing a significant transformation. Many factors have become the motivation for taking interest in cloud computing and these factors are the low cost hardware, storage capacity, increase in computing power and the tremendous growth in data size generating per day. The main challenge in the cloud is how to effectively store, query, analyze, and utilize immense datasets [8]. Some activities requires specialized models in the cloud computing domain in that for processing large datasets in clusters of computers, MapReduce model is used. MapReduce model comes under platform-as-a-service which offers high level of abstraction to make cloud programmable.

Google has successfully implemented data intensive paradigm called MapReduce to solve many large-scale computing problems [1]. The novel way to construct distributed applications for the cloud is to use MapReduce programming model. Its well known open-source implementation is Hadoop which is developed by Doug Cutting. The core of Hadoop includes file System, RPC, and serialization libraries and also it provides the basic services for building a cloud computing environment with commodity hardware. The two fundamental services of Hadoop are Hadoop Distributed File System (HDFS) [3] and MapReduce framework [1][5]. The Hadoop Distributed File System is a distributed file system which stores terabytes or petabytes of data; also it provides high speed access to the application data. It is highly designed to run on clusters of commodity machines. MapReduce framework is for processing large datasets on compute clusters in distributed way. MapReduce framework handles all complexities and distribution of the data as well as of map and reduce task.

The main point in using Hadoop is to handle large datasets efficiently. In this system, we have applied Hadoop MapReduce model to analyze web application log files. Log files are generated at a record rate as people use these web applications available in different areas such as shopping, banking, etc. Log file is a record of list of actions that have been occurred and also it keeps information about everything that goes in and out of the web server. Log files contain tons of information which is useful for making business decisions and future assessment. In order to analyze customer's behavior, market values for business, how our website is working, we need to process log files. Log file generation rate is nearly in some hundreds of TB's per day. Such massive amount of log data is difficult to store, analyze and utilize. Hadoop is the best fit in the cloud which can store such large log files and analyze them.

Proposed system uses Hadoop distributed file system to store log file and MapReduce programming model is used to write application for analyzing log file. First, log file is distributed over the nodes in a cluster and MapReduce is applied over them to get the analyzed results. The framework requires user to define two functions, Map and Reduce. MapReduce operates on each record in the log file and generates (key, value) pair as output where key is the field in the log file and value is the hit count for that particular field. Parallelization of MapReduce tasks makes execution faster. Pig queries aggregates MapReduce output from all the nodes in a cluster and then categorizes results according to the different fields in the log file [4].

All the details of the system are given in upcoming sections. Background details are presented in the next section. Section 3 provides complete view of proposed system and implementation details and results are shown in Section 4. Section 5 presents the conclusions.

II. LITERATURE SURVEY

Cloud computing is the enhanced version of grid or cluster. One of the nitty-gritty of cloud is that it can process with the huge amount of data within fraction of time as compared with the existing processing models. The exponential growth of data has made the world to live into the data age. Many tycoon companies like Yahoo, Google, Micrososft were finding difficulties in handling massive datasets. In 2004, Google introduced MapReduce and Google File System to scale up the data processing needs. Google's MapReduce is then implemented by many search engines and ultimately adopted by Hadoop. Now, Hadoop has become the core part of computing in many web companies [5][23]. Hadoop can handle terabytes or petabytes of data thus it is called as a Big Data technology.

In today's scenario, everything is going online which results into generating log files very fast. The way logs stores important information about customer's behavior and business, companies started storing their log data on a priority basis. Thus, log data has become big data [15]. In order to get more and more customers, they need to analyze which prior customers are interested, where it is more popular, which kind of service people are interested in, etc. Such kind of analysis helps in improving advertize of less popular services, promote popular services in order to make business scale. To analyze historical log files to make sense of the business for whole year, we need to store these log files for that we need reliable storage system. Thus, to solve problem of storing and analyzing large amount of log data, combined solution is needed. Hadoop is a good platform for storing and analyzing these huge log data. Two major components of Hadoop which are Hadoop Distributed File System and MapReduce take care of storage and large-scale data processing. Hadoop distributed file system stores petabytes of data by partitioning the data into small blocks and distributing them over multiple nodes in a cluster [3]. Map Reduce allows the distributed processing of the input log data in PetaBytes and could be able to come up with the result in amazingly less time.

*A. Difficulties in Existing Systems*

The term Big Data with MapReduce fascinated the big shot companies in the world like Yahoo, Facebook, Cloudera etc. and finally deflected the Moore's law little bit towards the saturation in the clock speed [5]. Traditionally, primary goal was to increase computing power of single machine. Then evolved distributed system which allowed a single job to run on multiple machines. From many years, High Performance Computing and Grid Computing have been doing data processing using SAN. The main disadvantage of using SAN is the single point of failure. Also at computation time, it copies data to compute node which is well suited for small amount of data only. For massively growing data, moving data to compute nodes is not a good idea. Hadoop MapReduce provides the data locality facility of moving computation to data rather moving data to computation which makes access to data fast [1][5]. And this is the reason for good performance of Hadoop MapReduce. Message Passing Interface used in previous technologies requires user to handle data flow explicitly wherein there is Application Programming Interface provided by the Hadoop works implicitly and embeds the business logic in the Mapper and Reducer Class [1]. Hadoop also provides cost effective storage as it runs over the commodity hardware.

Hadoop MapReduce also differs from traditional relational databases the many ways. Relational databases can handle only some gigabytes of data where Hadoop can handle terabytes or petabytes of data. Secondly, relational databases works over structured data only, there is a static schema [5][9][12]. Hadoop MapReduce is suitable for unstructured data such as text file as well as for semi structured data because it interpret the data at processing time.

III. PROPOSED WORK

*A. The Workflow of The System*

Our proposed system is composed of two phases involving log preprocessing and analysis phase. Block diagram of the system is shown in Figure 1. Input to the system is web application log file of banking server. Log file is a simple text file consisting of URL, date, hit, age, country, state and city. Preprocessing phase involves separation of fields using separator '#' and removing unwanted noisy data which could be multimedia files, style sheets, etc. Log file after preprocessing phase is stored into HDFS. Second phase is the analysis phase which involves distribution of the log file over Hadoop cluster, use of MapReduce algorithm and execution of pig query to integrate and categorize the analyzed results. Each task in analysis phase is explained below:

*1) Distribution of log file:* Hadoop breaks down input file into smaller blocks of equal size and distributes these blocks over multiple nodes in the Hadoop cluster. MapReduce processes these blocks in a parallel way.

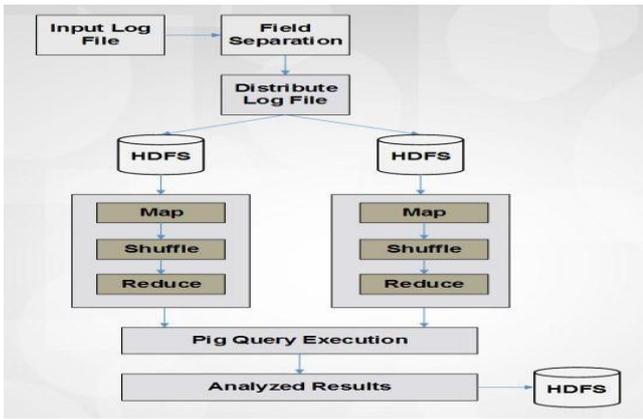

Fig. 1 Block Diagram

*2) MapReduce Algorithm:* MapReduce is a simple programming model which is easily scalable over multiple nodes in a Hadoop cluster. MapReduce job is written in Java consisting of Map and Reduce function. MapReduce takes log file as an input and feeds each record in the log file to the Mapper. Mapper processes all the records in the log file and Reducer processes all the outputs from the Mapper and gives final reduced results. Simply, Mapper filters and transforms the input into something that the Reducer can aggregate over. MapReduce execution steps are shown in Figure 2.

*a) Map Function:* Input to the map method is the InputSplit of log file. It produces intermediate results in (key, value) pairs. For each occurrence of key it emits (key, '1') pair. If there are n occurrences of key, then it produces n (key, '1') pairs. OutputCollector is the utility provided by MapReduce framework to collect output from mapper and reducer and reporter is to report a progress of application.

Map(LongWritable key, Text value, OutputCollector output, Reporter reporter)
{
    For each key in the value;
    EmitIntermediate(key, '1');
}

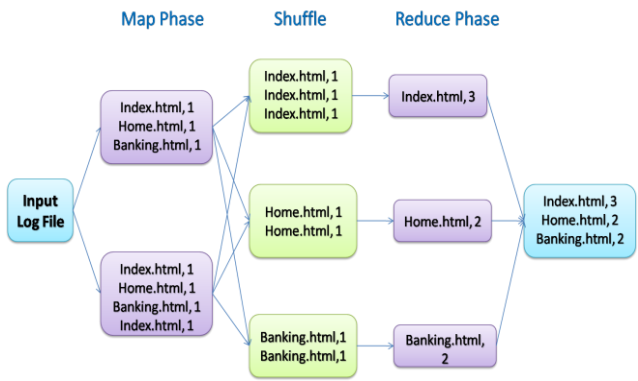

Fig. 2 MapReduce Framework

*b) Reduce Function:* Input to reduce method is (key, values) pairs. It sums together all counts emitted by map method. If input to the reduce method is (key, (1,1,1,....n times)) then it aggregates all the values for that key producing output (key, n) pair. OutputCollector and Reporter works in similar way as in map method.

reduce(Text key, Iterator values, OutputCollector output, Reporter reporter)
{
    int sum = 0;
    for each v in values;
    sum += ParseInt(v);
    output.collect(key,(sum));
}

*3) Pig Query:* Pig queries are written in Pig Latin language. Pig Latin statements are generally organized in the following manner:
- A LOAD statement reads data from the Hadoop file system.
- A series of "transformation" statements process the data.
- A STORE statement writes output to the Hadoop file system.

The example of Pig Latin program which operates on the results of MapReduce to get results of total hits per pages is given below:

A = load '/home/hadoop/output/1/part-00000' using PigStorage('\t') AS (page:chararray,hits:int);
B = load '/home/hadoop/output/2/part-00000' using PigStorage('\t') AS  (page:chararray,hits:int);
X = UNION A, B;
Y = FILTER X BY (page matches '^HitsPage-.*') ;
X = FOREACH Y GENERATE page,hits;
X =GROUP X by page;
X = FOREACH X GENERATE group , SUM(X.hits);
store X into 'Data/HitsPages' using PigStorage('\t','-schema');

*B. Three-Tier Architecture of The System*

The architecture of the proposed system is a three-tier architecture consisting of user interface, application code and data storage which is shown in Figure 3. The need of the three tier architecture is as it is the system which consists of distributed client - server design. User interface provides a way to the client to interact with the system. Application code is the middle tier between user interface and data store which performs operations on the client's request and provides results of user query. Third is the data layer which stores business data required by the application.

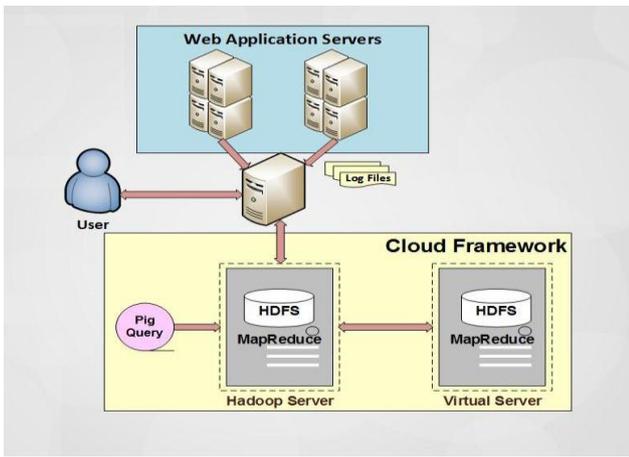

Fig. 3 System Architecture

Architecture is divided into three components as follows:

- User Interface: User has provided with the graphical user interface in which he has given provision of running web analytics over single-node cluster or multi-node cluster. User can distribute log file over multi node cluster and results of web analytics over that log file are provided to the user in the form of pie charts and bar charts. So user can check the analyzed results by viewing graphs implemented based on keys in the log file. User can be the web designer, marketing person or administrator. Through the analyzed results they will come to know what changes needs to do in the website or how to advertize and promote low priority services and how to increase performance.
- Business Logic: It is the middle tier which executes user's request by performing operations on the data stored in data store. In this system, user invokes to distribute log file on modes in a Hadoop cluster and run MapReduce in a distributed way. So the logic for these users's request is provided in this tier. So it distributes log files by dividing them in equal sized blocks over multiple node in a Hadoop cluster, performs Map and Reduce operations over blocks of log file to generate results depending on the key attributes in the log file. Also it does pig query execution over this results to separate out results according to key attributes to generate pie charts and bar charts for them. So when any change occurs we need to change business logic accordingly.
- Data Store: This layer provides data availability. Log file is the required data to process in this system. This log file is stored in HDFS by distributing it over a Hadoop cluster. HDFS works well for text files so it is a good fit to store log files in the simple text format.

## IV. IMPLEMENTATION

### A. Experimantal Setup

Log file used in our experiment is application server log file in simple text format. Experimental log file contains 100,000 records in it with each log having different fields as URL, date, hit, age, country, state, city. Log file is first preprocessed separating each field in it using separator '#'. Preprocessed log file is shown in Figure 4.

Hadoop is deployed on two machines creating multi node cluster. One server acts as a cloud server where actual log file is stored and other server works virtually. For parallel execution of MapReduce program log file is distributed evenly on both the nodes, each node having log file with 50,000 records in it. Running MapReduce job on both the nodes alongside, execution time is reduced. For optimizing results, Pig queries are written over the results of MapReduce job. All the experiments are implemented in Java. User can access the system via Internet in which facility is provided to distribute log file and to run MapReduce job over single node cluster or multi node cluster, and results of analysis are shown in the form of pie charts and bar charts.

```
pizza/index.html#13/01/2012#1#48#india#mh#pune
pizza/index.html#23/05/2012#1#37#india#mh#pune
/pizza/anywhere-banking.html#04/09/2012#1#39#india#mh#pune
/pizza/anywhere-banking.html#16/08/2012#1#32#india#mh#nashik
/pizza/cosmos-e-solutions-pvt-ltd.htm#03/10/2012#1#43#india#mh#bombay
/pizza/cosmos-e-solutions-pvt-ltd.htm#25/03/2012#1#40#india#mh#bombay
pizza/index.html#27/07/2012#1#62#india#mh#nashik
pizza/index.html#14/08/2012#1#31#india#mh#pune
pizza/index.html#14/11/2012#1#10#india#mh#bombay
/pizza/anywhere-banking.html#26/10/2012#1#38#india#mh#nashik
pizza/index.html#27/09/2012#1#35#india#mh#pune
pizza/index.html#11/06/2012#1#7#india#mh#pune
/pizza/cosmos-e-solutions-pvt-ltd.htm#29/04/2012#1#46#india#mh#bombay
pizza/index.html#21/03/2012#1#3#india#mh#nashik
/pizza/anywhere-banking.html#20/03/2012#1#2#india#mh#nashik
pizza/index.html#03/10/2012#1#23#india#mh#bombay
pizza/index.html#14/09/2012#1#41#india#mh#pune
pizza/index.html#19/05/2012#1#30#india#mh#pune
```

Fig. 4 Preprocessed Log File

### B. Results

Distribution of log file and MapReduce execution is shown in Figure 5 which also shows total time of execution of MapReduce job. Figure 6 shows results of analysis with provision of different charts. Results are provided according to field name in log file also it provides user to view hit count of that particular field in the form of pie chart or bar chart. As there are many fields in log file such as URL, date, hit, age, country, state, city, few outputs are shown in following figures. Figure 7 shows bar chart showing total hits for each city, total hits for quarter of the year in the form of pie chart is shown in Figure 8 and bar chart showing total hits for each page of website is shown in Figure 9.

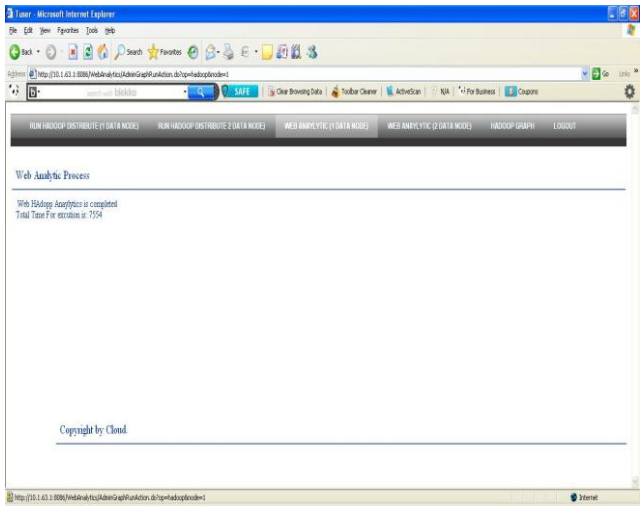

Fig. 5 MapReduce Execution

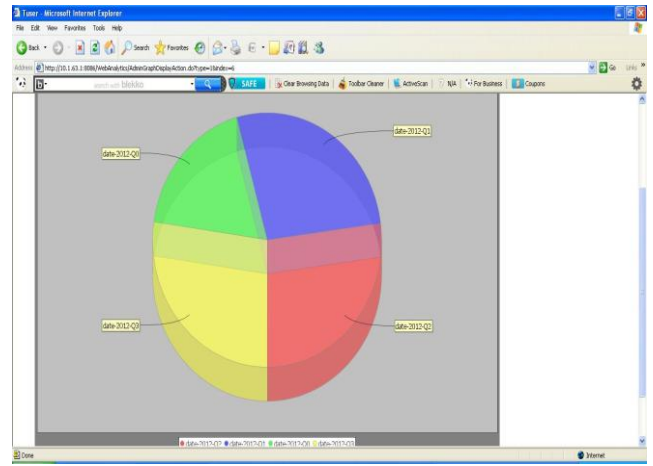

Fig. 8 Hits for Each Quarter of the Year (Pie Chart)

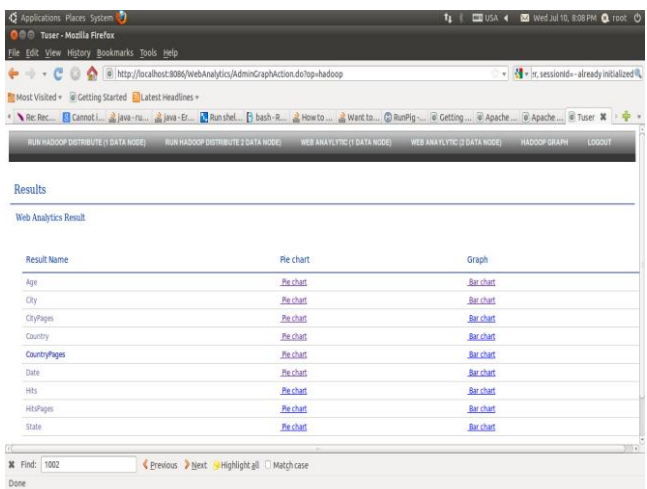

Fig. 6 Results of analysis

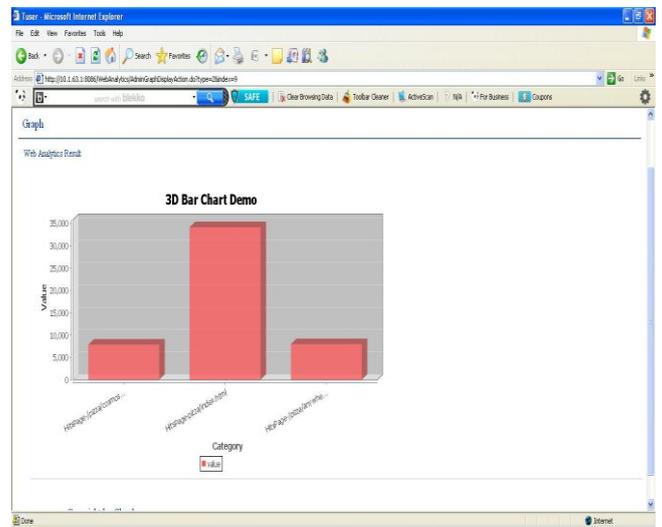

Fig. 9 Hits per Page (Bar Chart)

### C. Performance Analysis

Performance parameters of the system are number of records in the log file, number of nodes in the cluster and time required to process these records in the log file depending on the nodes in a cluster. Performance analysis is shown in Figure 11 in which performance graph is plotted against time required to process numbers of records on single node Hadoop cluster as well as multi node Hadoop cluster. Log file consists of 100,000 records so performance is evaluated for 20000, 40000, 60000, 80000 and 100000 records in the log file. Figure 10 shows execution details of different number of records on different clusters. Line graph in Figure 11 shows time required for multi node cluster is less than single node cluster as in multi node cluster tasks are performed in parallel manner so it takes less time to analyze given number of records. In single node cluster single machine performs task of analyzing records so MapReduce also requires time to execute given job.

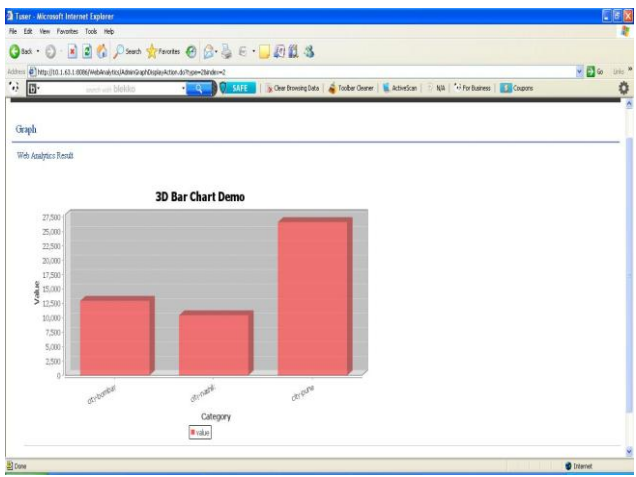

Fig. 7 Hits for Each City (Bar Chart)

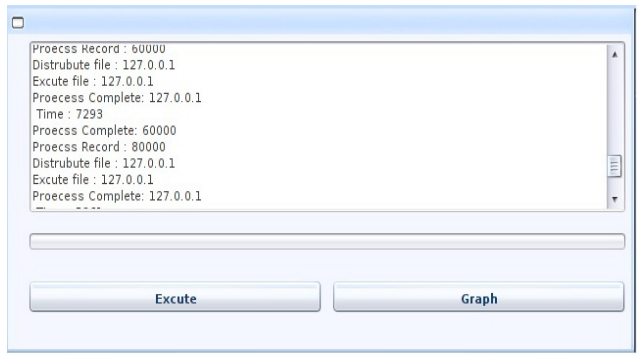

Fig. 10 Execution Process for Checking Performance

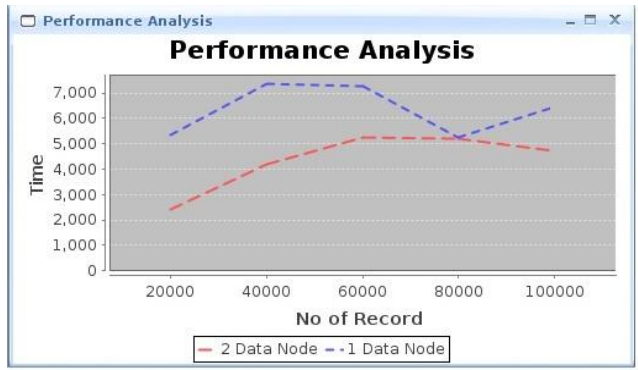

Fig. 11 Performance of Different Clusters

## V. CONCLUSIONS

We have presented best fit Hadoop MapReduce programming model for analyzing web application log files in cloud computing environment. In this system, data storage is provided using HDFS and MapReduce model applied over log files gives analyzed results in minimal response time. To get categorized results of analysis pig query is written over MapReduce result. Statistical record of analysis is shown in various charts as bar chart and pie chart which gives hit count for various parameters in log file. We have tested performance of the system against number of records, number of nodes in the cluster and experimental results show that as the number of nodes in cluster increases performance of the system also increases.